\documentstyle[preprint,aps]{revtex}
\begin{document}
\title{Simultaneous
Diagonal and Off Diagonal Order in the Bose--Hubbard Hamiltonian}
\author{R.T. Scalettar$^{\,1}$, G.G. Batrouni$^{2}$, A.P. Kampf$^{\,3}$,
G.T. Zimanyi$^{\,1}$}
\address{$^{1}$
Physics Department,
University of California,
Davis, CA 95616}
\address{$^{2}$
Groupe Matiere Condensee et Materiaux (URA 804),
Universite de Rennes I,
35042 Rennes CEDEX, France}
\address{$^{3}$
Institut f\"ur Theoretische Physik,
Universit\"at zu K\"oln,
50937 K\"oln,
Germany}
\date{\today}
\maketitle
\begin{abstract}
The Bose-Hubbard model exhibits a rich phase diagram consisting both of
insulating regimes where diagonal long range (solid) order dominates
as well as conducting
regimes where off diagonal long range order (superfluidity)
is present.  In this paper we describe the results of Quantum Monte Carlo
calculations of the
phase diagram, both for the hard and soft core cases,
with a particular focus on the possibility of simultaneous
superfluid and solid order.
We also discuss the appearance of phase separation in the model.
The simulations are compared with
analytic calculations of the phase diagram and spin wave
dispersion.

\end{abstract}
\newpage

\section{Introduction}

A lot of attention has been focussed on the interacting electron problem in the
last several decades, whereas the interacting boson problem has been considered
more often in the framework of specific applications only. However, there are
a number of important situations where the elementary excitations are either
intrinsically bosonic in character or else can usefully be viewed in terms of
bosonic models. $^{4}He$ is an example of the former situation,\cite{FISHER1}
while quantum spin systems,\cite{BSPIN} granular superconductors,\cite{BSUPER}
and flux lines in type--II superconductors\cite{BFLUX} are examples of the
latter.  Therefore it is important to understand in detail
the features of model boson systems, in much the same way that one studies
the Hubbard, Anderson, and t-J Hamiltonians for correlated fermions.
In this paper we consider a lattice model of interacting bosons, the
Bose Hubbard (BH) Hamiltonian:
\begin{equation}
H=-t\sum_{\langle ij\rangle}(a_{i}^{\dagger}a_{j}+a_{j}^{\dagger}a_{i})-\mu
\sum_{i}n_{i}+V_{0}\sum_{i}n_{i}^{2}+V_{1}\sum_{\langle ij\rangle}n_{i}n_{j}+
V_{2}\sum_{\langle\langle ik\rangle\rangle}n_{i}n_{k}\, .
\label{eq:eq1}
\end{equation}
Here $a_{i}$ is a boson annihilation operator at site $i$, and $n_{i}=
a_{i}^{\dagger}a_{i}$. The transfer integral $t=1$ sets the scale of the
energy, and $\mu$ is the chemical potential. $V_{0}, V_{1}$, and $V_{2}$ are
on--site, near--neighbor, and next--near--neighbor boson--boson repulsions.

The interactions $V_{0}, V_{1}$, and $V_{2}$ promote the formation of ``solid''
order, where the boson occupations fall into regular patterns, at special
densities commensurate with the lattice. The hopping matrix element $t$ favors
mobile bosons, and consequently a superfluid phase at $T=0$. In what follows
the nature of the correlation functions will be studied as we change the
Hamiltonian parameters and the density $\rho={1\over N}\sum_i\langle n_i
\rangle$.

When $V_{0}=\infty$, the BH model maps onto the quantum spin--1/2 Hamiltonian
\begin{equation}
H=-t\sum_{\langle ij \rangle}(S_{i}^{+}S_{j}^{-}+S_{j}^{+}S_{i}^{-})+V_{1}
\sum_{\langle ij\rangle}S_{i}^{z}S_{j}^{z}+V_{2}\sum_{\langle\langle ik\rangle
\rangle}S_{i}^{z}S_{k}^{z}-H_{z}\sum_{i}S_{i}^{z}.
\label{eq:eq2}
\end{equation}
The field $H_{z}=\mu-2V_{1}-2V_{2}$. Since $n_{i}\leftrightarrow S_{i}^{z}+
\frac12$, ordering of the density corresponds to finite wave vector Ising type
order. Similarly, $a_{i} \leftrightarrow S_{i}^{-}$ so that superfluidity maps
to ferromagnetic ordering in the XY plane. One of the things we shall be
interested in in this work is the possibility that density and
superfluid order are not
mutually exclusive. Indeed, at $V_{2}=H_{z}=0$, the special point $V_{1}=2t$
corresponds to the Heisenberg Hamiltonian where Ising and XY order coexist. It
has been suggested by various authors\cite{FISHERLIU,MATSUDA} that the addition
of further terms like $V_{2}$ or $H_{z}$ could stabilize this ``supersolid''
from a special symmetry point to a broader area of the phase diagram. Precisely
at the Heisenberg antiferromagnet (AF), the effect of a field $H_{z}$ is known:
It breaks the full rotational symmetry and selects ordering in the XY plane
since the spins can more easily take advantage of the field energy. This
argument has been used to suggest why doping favors the superconducting over
the CDW state in the negative--$U$ Hubbard model\cite{NEGU} where an analogous
``supersolid'' symmetry exists at half--filling.

While there have been many mean field (MF) studies of the spin Hamiltonian
Eq.~2, there have been to date only a few numerical studies
\cite{LOH,OTTERLO,BATROUNI1}. Monte Carlo studies of interacting quantum boson
and spin models provide a useful, exact method to study the nature of the
correlations on finite lattices. Combined with finite size scaling methods,
they can be used to extract information concerning the thermodynamic limit.
Boson simulations are somewhat easier than related path integral methods for
interacting electron systems, since they can utilize algorithms which scale
linearly with the lattice size and can reach essentially arbitrarily low
temperatures.

This paper is organized as follows: In Section II and III we determine
analytically the MF phase diagram, extending past work by considering
additional types of order, and describe spin wave
calculations of the dispersion relations in the various phases. In Section IV
we provide numerical results for the soft core model, extending our earlier
studies\cite{BATROUNI1}.
In Section V we describe new results for the hard--core
phase diagram. Conclusions are presented in Section VI.

\section{Mean Field Phase diagram}

Previous work established the MF phase diagram of the spin Hamiltonian
considering only the possibility of superfluidity and N\'eel--type ordering of
the density.\cite{FISHERLIU,MATSUDA,CHESTER} At half--filling, or
equivalently at zero magnetization $M_{z}=0$, for $V_1>2$ and $0<V_2<V_1-2$ the
spins form a N\'eel state, corresponding to a checkerboard Bose {\it solid}
with an ordering vector ${\bf k}_{*}=(\pi,\pi)$. For $V_2 > max~(V_1-2;0)$ a
ferromagnetic phase is formed, with a net moment $M_{xy}\ne 0$ and
$M_{z}\ne 0$. This phase corresponds to a {\it superfluid}, and is also stable
for arbitrary $V_1$ and $V_2$ away from half--filling. A fully polarized
magnetic phase in a strong magnetic field $H_z$, where only $M_{z}\ne 0$,
corresponds to a {\it Mott--insulator} with precisely one boson per site. As
the solid and the superfluid phases possess different broken symmetries, one
could expect that the transition between them is first order. However, a rather
different scenario has also been put forward, suggesting that the -- presumably
-- first order transition is split up into two distinct second order
transitions, where the two order parameters vanish at separate
points\cite{CHESTER,ANDREEV1,LEGGETT} In the regime between the two
transitions {\it both} order parameters are non--zero, hence it has been termed
a {\it supersolid}.\cite{FISHERLIU,MATSUDA,CHESTER} This intriguing possibility
is the subject of the investigations reported in this paper.

The mean field analysis indeed finds such a supersolid phase
\cite{FISHERLIU,MATSUDA,CHESTER}, although in the hard core limit longer range
forces ($V_{2}> 0$) are needed to stabilize it. However, recently it was
claimed that this conclusion changes in the soft core case, and a supersolid
phase exists with nearest neighbor interaction alone\cite{STROUD}. Finally,
recent studies on the related Heisenberg model with competing first and second
neighbor couplings $J_{1}$ and $J_{2}$ established the possibility of
additional phases: a collinear phase, with alternating lines of up and down
spins, at large $J_{2}/J_{1}$\cite{singh,dagotto,larkin}, and a disordered
phase at intermediate values of $J_{2}/J_{1}$. \cite{sachdev,kivelson} These
differing results clearly call for a reinvestigation of the problem.

The MF phase diagram of the spin Hamiltonian Eq.~2
worked out by Matsuda and Tsuneto\cite{MATSUDA}, and described above,
allowed only for a two--sublattice
magnetic ordering of the spins corresponding to a N\'eel solid.
Representing the spins by classical vectors of
length $S$ we extend earlier MF analyses\cite{MATSUDA,BRUDER} for the
case of a square lattice to include also the possibility of a collinear phase
which is expected to form for intermediate to large next--near--neighbor
repulsion $V_2$ (see Fig.~1). Assuming that the spins are ordered in the XZ
plane, the MF energies per spin, $e_{N}$ and $e_{C}$, of the N\'eel
and collinear spin configurations are given by
\begin{eqnarray}
e_{N}&=&-4S^2\sin\theta_A\sin\theta_B+2S^2V_1\cos\theta_A\cos\theta_B+V_2S^2
\left(\cos^2\theta_A+\cos^2\theta_B\right)- \nonumber \\
&&-{H_z\over 2}S\Big(\cos\theta_A+\cos\theta_B\Big) \\
e_{C}&=&-S^2\left(\sin\theta_{R1}+\sin\theta_{R2}\right)^2+{V_1S^2
\over 2}\left(\cos\theta_{R1}+\cos\theta_{R2}\right)^2+2V_2S^2\cos\theta_{R1}
\cos\theta_{R2}- \nonumber \\
&&-{H_z\over 2}S\left(\cos\theta_{R1}+\cos\theta_{R2}\right)\, .
\label{eq:eq3,4}
\end{eqnarray}
$\theta_A$ and $\theta_B$ are the angles between the spin direction and the
$z$--axis on sublattice $A$ and $B$, respectively. $\theta_{R1}$ and
$\theta_{R2}$ are the corresponding angles in the collinear phase on even and
odd rows. The different phases are identified as follows:
\begin{eqnarray}
\cos\theta_{A}=\cos\theta_{B}<1~~{\rm or}~~\cos\theta_{R1}=\cos\theta_{R2} &<&
1~~\hskip0.4cm {\rm Superfluid}\nonumber\\
\cos\theta_{A}=-\cos\theta_{B}&=&1 ~~\hskip0.4cm{\rm Neel ~~Solid}\nonumber\\
\cos\theta_{R1}=-\cos\theta_{R2}&=&1~~\hskip0.4cm {\rm Collinear ~~Solid}
\nonumber\\
\sin\theta_{A}\ne\sin\theta_{B}~~{\rm and}~~-1<\cos\theta_{A}\ne-\cos\theta_{B}
&<&1 ~~\hskip0.4cm {\rm Neel ~~Supersolid}\nonumber\\
\sin\theta_{R1}\ne\sin\theta_{R2}~~{\rm and}~~-1<\cos\theta_{R1}\ne -\cos
\theta_{R2}&<&1 ~~\hskip0.4cm{\rm Collinear ~~Supersolid}\nonumber\\
\cos\theta_{A}=\cos\theta_{B}=1 ~~{\rm or}~~
\cos\theta_{R1}=\cos\theta_{R2}&=&1 ~~\hskip0.4cm{\rm Mott ~~phase}
\label{eq:eq5}
\end{eqnarray}

We performed the MF analysis in the same spirit as in
Refs.\cite{FISHERLIU,MATSUDA,CHESTER}. One proceeds by minimizing $e_{N}$ and
$e_{C}$ separately with respect to the angles $\theta_A,\,\theta_B$ and
$\theta_{R1},\,\theta_{R2}$, respectively. Then the results for fixed magnetic
field $H_z$ are translated to fixed magnetisation, i.e. boson density. Finally,
we compare the energies of the different phases to obtain the complete MF phase
diagram of the spin Hamiltonian Eq.~2. Explicitly, for
two--sublattice N\'eel type ordering we find the following phases for $V_1>2$
and $0<V_2<V_1-2$:
\begin{eqnarray}
&m&=0 \hskip5.0cm {\rm Solid} \nonumber \\
0<&m&<{1\over 2}\sqrt{{V_1-V_2-2\over V_1-V_2+2}}\hskip 2.5cm
{\rm Neel~~ Supersolid} \nonumber \\
{1\over 2}\sqrt{{V_1-V_2-2\over V_1-V_2+2}}<&m&<{1\over 2}\hskip5.3cm
{\rm Superfluid}\nonumber \\
&m&={1\over 2}\hskip5.0cm {\rm Mott~~ Insulator}
\label{eq:eq6}
\end{eqnarray}
where $m=|\rho-{1\over 2}|$ is the magnetisation of the system. For $0<V_1<2$
there is no N\'eel order and for $m\ne 0$ the MF ground state is always
superfluid.

Similarly we analyze the phase diagram following from minimizing $e_{C}$ for
the ordered collinear spin structures corresponding to an ordering wave vector
${\bf k}_{*}=(0,\pi)$ or $(\pi,0)$. At half--filling the collinear solid (see
Fig.~1) is realized for arbitrary values of the near--neighbor repulsion $V_1$.
The reason is that at half--filling the energy for the collinear solid is
$e_{C}=-V_2/2$, i.e. independent of $V_1$ due to the cancellation of $S_i^z
S_j^z$ energies for near--neighbor sites on the same and neighboring rows. Away
from half--filling only the superfluid minimizes $e_{C}$ for $V_2<2$. For
$V_2>2$ a collinear supersolid appears in the phase diagram and the boundary
between the superfluid and the collinear supersolid is determined by
\begin{eqnarray}
0<&m&<{1\over 2}\sqrt{{V_2-2\over V_2}}\hskip 3.2cm{\rm Collinear\,\,
Supersolid} \nonumber \\
{1\over 2}\sqrt{{V_2-2\over V_2}}<&m&<{1\over 2}\hskip4.7cm{\rm Superfluid}
\label{eq:eq7}
\end{eqnarray}
which is again independent of $V_1$. For $V_2>2$ the collinear supersolid phase
occurs in a density strip of width $\sqrt{(V_2-2)/V_2}$ around half--filling.

Given the MF solution for $e_{N}$ and $e_{C}$ separately, a comparison
for the energies of the different phases allows to map out the complete
mean--field phase diagram of the spin Hamiltonian Eq.~2. E.g. at half--filling,
$m=0$, we have to compare
\begin{eqnarray}
e_{C}&=&-{1\over 2}V_2 \hskip4.1cm{\rm Collinear\,\, Solid} \nonumber \\
e_{N}&=&e_C=-1\hskip3.8cm {\rm Superfluid} \nonumber \\
e_{N}&=&{1\over 2}(V_2-V_1)\hskip3.5cm {\rm Neel\,\, Solid}
\label{eq:eq8}
\end{eqnarray}
The resulting phase diagram is shown in Fig.~2. Interestingly, for $2<V_1<4$
increasing $V_2$ drives two transitions: first increasing $V_2$ frustrates the
N\'eel solid and leads to a transition to a superfluid. Increasing $V_2$
further stabilizes collinear order and leads to a transition from a superfluid
to a collinear solid.\cite{NIYAZ}

Away from half--filling, $0<m<1/2$, no solids, neither N\'eel nor collinear are
MF solutions. Instead, transitions occur between the superfluid, and
the N\'eel and collinear {\it supersolid} phases. The boundaries between the
different phases are given by
\begin{eqnarray}
V_2&=&V_1-2{1+4m^2\over 1-4m^2}\hskip 2.4cm{\rm Superfluid\,\, to\,\, Neel\,\,
Supersolid}
\nonumber \\
V_2&=&{2\over 1-4m^2} \hskip 3.4cm{\rm Superfluid\,\, to\,\, Collinear\,\,
Supersolid}\nonumber \\
V_2&=&{1\over 2}V_1-{4m^2\over 1-4m^2} \hskip 2.1cm{\rm Neel\,\, to\,\,
Collinear\,\, Supersolid}
\label{eq:eq9}
\end{eqnarray}
Finite doping leads to a rigid shift of the phase boundary lines
obtained at half--filling with the solid replaced by supersolid phases. For
\begin{equation}
2{1+4m^2\over 1-4m^2}<V_1<4{1+2m^2\over 1-4m^2}
\label{eq:eq10}
\end{equation}
this still allows for two transitions with increasing $V_2$, from a N\'eel
supersolid to a superfluid to a collinear supersolid. The $V_1$--$V_2$ phase
diagram for a fixed magnetisation $m=0.2$ is shown in Fig.~3. In addition,
Figs.~4 and 5 show the phase boundaries in the $V_2$, $m$ plane for a fixed
value of $V_1$ and in the $V_1$, $m$ plane for a fixed value of $V_2$,
respectively.

Recently it was claimed that a finite core repulsion $V_0<\infty$
qualitatively changes this picture.\cite{STROUD} Supersolids were found to
exist even {\it at half--filling}, moreover {\it without the next nearest
neighbor repulsion $V_{2}$}. To study these claims we extend the MF
analysis by introducing an approximate soft core representation allowing the
spin length $S$ to be a variational parameter and adding a term
$H_{constraint}=V_{0}\sum_{i}({\bf S}_{i}^2-1)^2$ to the Hamiltonian. The
minimization of the ground state energy is now done separately with respect to
$S_{x}^A, S_{x}^B$ and $S_{z}^A, S_{z}^B$.

We expand the ground state energy around the superfluid phase, and consider the
eigenvalues corresponding to small spatial modulations of the density and
superfluid order parameter, in effect generating a Ginzburg--Landau type
expression. The superfluid--collinear supersolid transition is studied by
writing
\begin{eqnarray}
S_z^A&=&m-\epsilon \hskip1.2cm S_z^B=m+\epsilon \nonumber\\
S_x^A&=&s-\delta \hskip1.3cm S_x^B=s+\delta
\label{eq:eq11}
\end{eqnarray}
and expanding to second order in the (small) fluctuations $\epsilon$ and
$\delta$. The expectation value $e$ of the ground state energy per site takes
the form
\begin{eqnarray}
e&=&e_{SF}-4V_2\epsilon^2+V_0\left[(12s^2+4m^2-1)\delta^2+(12m^2+4s^2-1)
\epsilon^2+16sm\epsilon\delta\right]\, ,\nonumber\\
e_{SF}&=&-8s^2+4(V_1+V_2)m^2+{1\over 8}V_0(4s^2+4m^2-1)^2 \, .
\label{eq:eq12}
\end{eqnarray}
The ground state energy is the sum of eigenvalues of a matrix in the
$(\epsilon,\delta)$ space. First we solve for $s$ at fixed number of particles,
i.e. fixed $m$, in the superconducting state where $\delta=\epsilon=0$, and
obtain
\begin{equation}
s^2={1\over 4}-m^2+{2\over V_0}\, .
\label{eq:eq13}
\end{equation}
A zero eigenvalue of the energy matrix signals the phase transition. The
condition for the vanishing of the determinant can be solved for $V_2$ for
arbitrary $m$
\begin{equation}
V_2=2+{8m^2\over 1-4m^2+12/V_0}\, ,
\label{eq:eq14}
\end{equation}
which gives the phase boundary between the superfluid and the collinear
supersolid. With the same procedure the phase boundary between the superfluid
and the N\'eel supersolid is at $V_{2}=V_{1}-2-16m^2/[1-4m^2+16/V_{0}]$.
As in the hard core case the phase diagram displays N\'eel-- and collinear
supersolid, and
superfluid phases. At half--filling the supersolid phases vanish, and two
insulating solids are direct neighbors to the superfluid, in contrast with the
result of Ref.~\cite{OTTERLO}.
This result is independent of $V_{0}$, {\it i.e.}
it is true both in the soft and hard core limits, in agreement with the above
hard core MF calculation.

\section{Spin Wave Analysis}

The analyses of the spin wave fluctuations which exist in the
literature\cite{FISHERLIU,CHESTER,CHENG} are in disagreement. The spectrum has
been found to be either linear\cite{FISHERLIU} or quadratic\cite{CHESTER,CHENG}
{\it at} the solid--supersolid phase boundary. This dependence is crucial for
numerical studies, as it determines the dynamical critical exponent $z$ and
thereby the appropriate finite size scaling of the lattice.

To settle the issue, we redo the linear spin wave theory analysis for the spin
model of Eq.~2 and determine the spectrum in the superfluid, the N\'eel solid
and the N\'eel supersolid. Again we assume that the spins are ordered in the XZ
plane with an angle $\theta_{A(B)}$ to the $z$--direction. On each sublattice
the spin quantisation axis is rotated to align the spins along the local
direction of the magnetisation by
\begin{equation}
{\bf S}_{i\in A(j\in B)}=\left[\matrix{\cos\theta_{A(B)}&0&-\sin\theta_{A(B)}
\cr 0&1&0\cr\sin\theta_{A(B)}&0&\cos\theta_{A(B)}\cr}\right]{\hat{\bf S}}_{i\in
A(j\in B)}\, .
\label{eq:eq15}
\end{equation}
To diagonalize the spin Hamiltonian Eq.~2 in terms of the rotated spins
${\hat{\bf S}}$ we introduce spin raising and lowering operators $\hat{a}^+$
and $\hat{a}$ on sublattice $A$ by
\begin{eqnarray}
\hat{S}^+_{i\in A}&=&\hat{S}^x_{i\in A}+i\hat{S}^y_{i\in A}=\hat{a}_i^+
\nonumber \\
\hat{S}^-_{i\in A}&=&\hat{S}^x_{i\in A}-i\hat{S}^y_{i\in A}=\hat{a}_i
\nonumber \\
\hat{S}^z_{i\in A}&=&{1\over 2}-\hat{a}_i^+\hat{a}_i
\label{eq:eq16}
\end{eqnarray}
which obey the usual bosonic commutation relations in the large $S$ limit.
Similarly, operators $\hat{b}^+$ and $\hat{b}$ are introduced on sublattice
$B$. After Fourier transformation this leads, up to a constant energy shift, to
the linear spin wave Hamiltonian
\begin{eqnarray}
H_{SW}=\sum_{\bf k}&{'}&\left[ H_{11}\left(\hat{a}_{\bf k}^+\hat{a}_{\bf k}+
\hat{a}_{-{\bf k}}^+\hat{a}_{-{\bf k}}\right)+H_{33}\left(\hat{b}_{\bf k}^+
\hat{b}_{\bf k}+\hat{b}_{-{\bf k}}^+\hat{b}_{-{\bf k}}\right)+H_{21}\left(
\hat{a}_{\bf k}^+\hat{a}_{-{\bf k}}^++\hat{a}_{\bf k}\hat{a}_{-{\bf k}}\right)+
\right. \nonumber \\
&&\left.+H_{34}\left(\hat{b}_{\bf k}^+\hat{b}_{-{\bf k}}^++\hat{b}_{\bf k}
\hat{b}_{-{\bf k}}\right)+H_{31}\left(\hat{a}_{\bf k}^+\hat{b}_{\bf k}+
\hat{a}_{-{\bf k}}^+\hat{b}_{-{\bf k}}+\hat{a}_{\bf k}\hat{b}_{\bf k}^++
\hat{a}_{-{\bf k}}\hat{b}_{-{\bf k}}^+\right)+ \right. \nonumber \\
&&\left.+H_{41}\left(\hat{a}_{\bf k}^+\hat{b}_{-{\bf k}}^++\hat{a}_{-{\bf k}}^+
\hat{b}_{\bf k}^++\hat{a}_{\bf k}\hat{b}_{-{\bf k}}+\hat{a}_{-{\bf k}}
\hat{b}_{\bf k}\right)\right]
\label{eq:eq17}
\end{eqnarray}
neglecting higher order terms in $\hat{a}$ and $\hat{b}$. Due to the
two--sublattice structure the ${\bf k}$--sum is restricted to half of the
Brillouin zone, i.e. to momenta with $\cos(k_x)+\cos(k_y)\ge 0$. In the
superfluid and the supersolid phase the ${\bf k}$--dependent coefficients in
Eq.~17 are given by:
\begin{eqnarray}
H_{11}=2{\sin\theta_B\over\sin\theta_A}+H_{21}\hskip0.3cm , \hskip0.3cm
H_{33}=2{\sin\theta_A\over\sin\theta_B}+H_{34}
\nonumber \\
\hskip0.5cm H_{21}={1\over 2}V_2\sin^2\theta_A\, \gamma_{\bf k}^{(2)}
\hskip0.3cm ,\hskip0.3cm H_{34}={1\over 2}V_2\sin^2\theta_B\, \gamma_{\bf k}^{
(2)}\nonumber \\
\hskip0.4cm H_{31}=\gamma_{\bf k}^{(1)}\left[-1-\cos\theta_A\cos
\theta_B+{1\over 2}V_1\sin\theta_A\sin\theta_B\right]\hskip0.3cm ,
\hskip0.3cm H_{41}=
H_{31}+2\gamma_{\bf k}^{(1)}
\label{eq:eq18}
\end{eqnarray}
where $\gamma_{\bf k}^{(1)}={1\over 2}(\cos(k_x)+\cos(k_y))$ and $\gamma_{\bf
k}^{(2)}=\cos(k_x)\cos(k_y)$. The coefficients of the first order terms are
required to vanish\cite{kanamori} which leads to the conditions
\begin{eqnarray}
{H_z\over 2}\sin\theta_A&=&2\cos\theta_A\sin\theta_B+V_1\cos\theta_B\sin
\theta_A+V_2\cos\theta_A\sin\theta_A\nonumber \\
{H_z\over 2}\sin\theta_B&=&2\cos\theta_B\sin\theta_A+V_1\cos\theta_A\sin
\theta_B+V_2\cos\theta_B\sin\theta_B \, .
\label{eq:eq19}
\end{eqnarray}
These two equations determine the angles $\theta_A$ and $\theta_B$ for a given
value of the magnetic field $H_z$. The solutions of Eqs.~19 determine the phase
diagram of the model. These equations fully coincide with the ones obtained by
minimizing the free energy in the previous section. We have already used Eq.~19
to eliminate the magnetic field in the expressions for the coefficients
$H_{11}$ and $H_{33}$ in Eq.~18 of the spin wave Hamiltonian. However, for the
N\'eel solid and the Mott insulator phase where both $\sin\theta_A=0$ and
$\sin\theta_B=0$ the elimination is not possible and instead $H_{11}$ and
$H_{33}$ are given by
\begin{eqnarray}
H_{11}&=&V_1-V_2-{1\over 2}H_z\hskip1.5cm H_{33}=V_1-V_2+{1\over 2}H_z\,
\hskip1cm {\rm Neel\,\, Solid}\nonumber\\
H_{11}&=&H_{33}=-V_1-V_2+{1\over 2}H_z\,\hskip4.6cm {\rm Mott\,\, Insulator}
\label{eq:eq20}
\end{eqnarray}

Eq.~17 is diagonalized by a generalized Bogoliubov transformation using the
equation of motion $i\partial_t\hat{a}_{\bf k}=[\hat{a}_{\bf k},H_{SW}]_-$ with
$\hat{a}_{\bf k}\propto e^{-i\omega_{\bf k}t}$. In the boson language the
Bogoliubov transformation involves coupled density and phase modes. As a result
we obtain the spin wave dispersion in the form
\begin{eqnarray}
\omega_\pm^2({\bf k})=&{1\over 2}&\left\lbrace H_{11}^2-H_{21}^2+2H_{31}^2-2
H_{41}^2+H_{33}^2-H_{34}^2\pm\Big\lgroup\left(H_{11}^2-H_{21}^2-H_{33}^2+
H_{34}^2\right)^2+ \right. \nonumber \\
&&\left.\hskip0.3cm+4\Big(\left[H_{11}-H_{21}\right]\left[H_{31}+H_{41}\right]+
\left[H_{33}+H_{34}\right]\left[H_{31}-H_{21}\right]\Big)\cdot \right.
\nonumber \\
&&\left.\hskip0.6cm\cdot\Big(\left[H_{33}-H_{34}\right]\left[H_{31}+H_{41}
\right]+\left[H_{11}+H_{21}\right]\left[H_{31}-H_{21}\right]\Big)\Big
\rgroup^{1/2}\right\rbrace
\label{eq:eq21}
\end{eqnarray}

Typical dispersions are shown in Fig.~6 in the different phases and on the
phase boundaries. We now overview the dispersion relations in the four phases.
\hfill\break
{\it i)} In the N\'eel solid which is realized for $V_2\le V_1-2$ and $H_z<2
\sqrt{(V_1-V_2)^2-4}$ the spin wave dispersion is given by
\begin{equation}
\omega_\pm({\bf k})=\sqrt{(V_1-V_2)^2-\left(2\gamma_{\bf k}^{(1)}\right)^2}\pm
{1\over 2}H_z\, .
\label{eq:eq22}
\end {equation}
Thus, there are two excitation branches in a halved magnetic Brillouin zone.
Both branches are gapped. \hfill\break
{\it ii)} In the superfluid there is a Goldstone mode of linear $k$
dependence at small $k$, and a well developed minimum around ${\bf k_*}=(\pi,
\pi)$. Taking the continuum limit carefully identifies this with the roton part
of the helium dispersion. Explicitly, with $s=\sin\theta_A=\sin\theta_B$, the
dispersion in the extended zone is given by
\begin{equation}
\omega^2({\bf k})=2\left(1-\gamma_{\bf k}^{(1)}\right)
\left(2(1-\gamma_{\bf k}^{(1)})+s^2\left[V_2\gamma_{\bf
k}^{(2)}+(2+V_1)\gamma_{\bf k}^{(1)}\right]\right).
\label{eq:eq23}
\end{equation}
\noindent
{\it iii)} In the Mott insulator phase for fields $H_z>2(2+V_1+V_2)$ all spins
are aligned along the magnetic field direction. There is a single gapped mode
in the extended $1^{st}$ Brillouin zone with the dispersion given by
\begin{equation}
\omega({\bf k})={H_z\over 2}-V_1-V_2-2\gamma_{\bf k}^{(1)}\, .
\label{eq:eq24}
\end{equation}
\noindent
{\it iv)} Finally, in the supersolid phase one has a gapless linear mode, and a
gapped one, again in the halved magnetic zone.

To clarify the physics of the transitions we concentrate on the dispersion at
${\bf k}\approx 0$ and ${\bf k}\approx (\pi,\pi)$ at the phase boundaries. At
the supersolid--N\'eel solid transition the critical mode is the Goldstone mode
at small $k$. At the critical magnetic field, $H_z^c=2\sqrt{(V_1-V_2)^2-4}$,
which determines the N\'eel solid to supersolid boundary by the vanishing of
the gap of the lower excitation branch of the solid, we perform the small $k$
expansion for $\omega_-({\bf k})$ from Eq.~22. For $V_1>V_2+2$ where the solid
exists at half--filling we obtain
\begin{equation}
\omega_-({\bf k})\approx {1\over 2}\,{1\over\sqrt{(V_1-V_2)^2-4}}~k^2\, .
\label{eq:eq25}
\end{equation}
This means that the linear mode of the supersolid softens into a quadratic one
at the boundary -- signalling the destruction of superfluidity -- before
lifting off into a gapped mode inside the solid phase. This yields a quantum
critical exponent $z=2$. This value of $z$ agrees with that of
Chester\cite{CHESTER} and Cheng\cite{CHENG}, but differs from that of Liu and
Fisher\cite{FISHERLIU}, who obtain $z=1$. We feel, however, that the softening
of the Goldstone mode is a physically realistic picture, supporting our result.

At the generic superfluid--to--N\'eel supersolid transition the critical mode
is at
${\bf k}_*=(\pi,\pi)$. Inside the superfluid phase the roton minimum is at this
wavevector. However in the solid, because of the zone--halving, this roton
minimum is folded back to ${\bf k}=0$. In the superfluid where $\sin\theta_{A}=
\sin\theta_{B}=s$ we study the small $k$ expansion of the single mode in the
neighborhood of ${\bf k}=(0,0)$ and ${\bf k}=(\pi,\pi)$. (Note that the N\'eel
supersolid is only realised for $V_2<V_1-2$.)
\begin{eqnarray}
\omega^2({\bf k})&\approx &{s^2\over 2}\left(2+V_2+V_1\right) ~k^2\nonumber\\
\omega^2((\pi,\pi)-{\bf k})&\approx &8\Delta^2+\left(2-3\Delta^2-V_2s^2\right)
\label{eq:eq26}
\end{eqnarray}
where $\Delta^2=2+(s^2/2)\left[V_2-2-V_1\right]$. At the boundary to the N\'eel
supersolid which is reached at a magnetic field $H_z=2\sqrt{(V_1-V_2)^2-4}(V_1+
V_2+2)/(V_1-V_2+2)$ the mean field conditions in Eq.~19 tell that exactly at
the transition the roton gap $\Delta$ disappears: {\it the solidification is
signalled by the softening--out of the roton mode of the superfluid}. The
dispersion relation of the rotons also changes from a quadratic to a linear
minimum, hence $z=1$.

Two remarks are in order here. First, recalling the original Landau argument
about superfluidity it is clear that a vanishing roton energy leads to a
vanishing critical velocity. So, while the superfluid order parameter remains
finite through the supersolid transition, the critical velocity collapses to
zero. {\it Inside} the supersolid phase it again assumes a finite value, as the
second excitation branch becomes gapped.

Second, one can raise the question of how this picture is going to be
modified in the absence of an underlying lattice. In this continuum limit the
modes which go soft are located at a finite magnitude of $k$, i.e. on a {\it
ring} in momentum space. This means that the phase space for these excitations
is much larger than for the usual Goldstone modes, which are centered around
$k\sim 0$. It then is possible that these excitations may give rise to a
fluctuation--induced first order transition instead of the second order one
taking place on the lattice\cite{KHMELNITSKI}.

Similar expansions can be used to study the case of half--filling. In this
particle--hole symmetric case, not surprisingly both transitions have
$z=1$. In a recent Monte Carlo study\cite{OTTERLO} the same $z$ value was used
in choosing the lattice size to study both the superfluid--supersolid and
supersolid--solid transitions, whereas we find $z=1$ and $z=2$, respectively
off half--filling.
It appears that our results call for the repetition of the numerical
simulations with different $z$ factors when $\rho \neq 1/2$.

Finally at high fields, at the superfluid--to--Mott insulator transition the
Goldstone mode softens out again, leading to $z=2$, in agreement with earlier
field theoretical predictions\cite{FISHER1} and numerical
simulations\cite{BATROUNI}.

We have repeated the spin wave calculation for the
collinear ordering for an ordering wave vector ${\bf k}_*=(0,\pi)$. In this
case the coefficients of the Hamiltonian Eq.~17 outside the collinear solid and
the Mott insulating phases are given by
\begin{eqnarray}
H_{11}=\left(1+{\sin\theta_{R2}\over\sin\theta_{R1}}\right)+{1\over
4}\left(V_1\sin^2\theta_{R1}-2-2\cos^2\theta_{R1}\right)\cos(k_x)\, \nonumber\\
H_{33}=\left(1+{\sin\theta_{R1}\over\sin\theta_{R2}}\right)+{1\over
4}\left(V_1\sin^2\theta_{R2}-2-2\cos^2\theta_{R2}\right)\cos(k_x)\, \nonumber\\
H_{21}={1\over 4}\sin^2\theta_{R1}\left(2+V_1\right)\cos(k_x)\hskip0.4cm ,
\hskip0.4cm H_{34}={1\over 4}\sin^2\theta_{R2}\left(2+V_1\right)\cos(k_x)\, ,
\nonumber \\
H_{31}={1\over 4}\left[-2-2\cos\theta_{R1}\cos\theta_{R2}+V_1\sin\theta_{R1}
\sin\theta_{R2}\right]\cos(k_y)+{V_2\over 2}\sin\theta_{R1}\sin\theta_{R2}\,
\gamma_{\bf k}^{(2)} \nonumber \\
H_{41}=H_{31}+\cos(k_y)\hskip4.3cm \, .
\label{eq:eq27}
\end{eqnarray}
The MF conditions are read off from the vanishing of the terms linear in the
spin wave operators as before
\begin{eqnarray}
H_z \sin\theta_{R1}&=&2\cos\theta_{R1}(\sin\theta_{R1}+\sin\theta_{R2})+V_1
\sin\theta_{R1}(\cos\theta_{R1}+\cos\theta_{R2})+2V_2\cos\theta_{R2}\sin
\theta_{R1}\nonumber \\
H_z\sin\theta_{R2}&=&2\cos\theta_{R2}(\sin\theta_{R1}+\sin\theta_{R2})+
V_1\sin\theta_{R2}(\cos\theta_{R1}+\cos\theta_{R2})+2V_2\cos\theta_{R1}\sin
\theta_{R2} \, .
\label{eq:eq28}
\end{eqnarray}
For the superfluid and the Mott insulating phase the spin wave dispersions are
obtained identical to the ones derived above in Eq.~23 and Eq.~24. In the
collinear solid with $\sin\theta_{R1}=\sin\theta_{R2}=0$ the coefficients
$H_{11}$ and $H_{33}$ are replaced by
\begin{equation}
H_{11}=V_2-{1\over 2}H_z-\cos(k_x)\hskip1.5cm H_{33}=V_2+{1\over 2}
H_z-\cos(k_x)\, .
\label{eq:eq29}
\end{equation}

The two gapped modes in the collinear solid for magnetic fields $H_z\le2\sqrt{(
V_2-1)^2-1}$ and $V_2\ge 2$ follow as
\begin{equation}
\omega_\pm ({\bf k})=\sqrt{\left(V_2-\cos(k_x)\right)^2-
\cos^2(k_y)}\pm {1\over 2}H_z\, .
\label{eq:eq30}
\end{equation}
As for the N\'eel supersolid, the collinear supersolid has one gapless
linear mode at small $k$ and a gapped one in the halved magnetic Brillouin zone
which in the case of collinear ordering with wave vector ${\bf k_*}=(0,\pi)$ is
determined by $|k_y|\le \pi/2$. The transition from the superfluid to the
collinear supersolid is now driven by the softening of the roton mode at
${\bf k_*}=(0,\pi)$. The dynamical exponent is again $z=1$. Also the exponents
at the superfluid to collinear solid and at the solid to supersolid transition
are identical to the exponents found for the N\'eel ordering transitions.

\section{Simulations of the Soft Core Model}

\noindent
{\bf Results at half--filling}

In this section we describe the results of numerical simulations,
and compare them with the picture gained from the analytical considerations.
Our Monte Carlo calculations are performed using a path integral representation
on the BH partition function by discretizing the
inverse temperature $\beta$ into $L_{\tau}$ intervals,
$\beta=L_{\tau}\Delta \tau$.  A description of the technical details
is contained in \cite{BATROUNI2}.
In order to characterize the phase diagram, we measure the boson winding number
to determine the superfluid density $\rho_{s}$. We also measure the
density--density correlations $c({\bf l})$ and their Fourier transform, the
structure factor $S({\bf k})$.
\begin{eqnarray}
c({\bf l})&=&\langle n({\bf j},\tau)n({\bf j}+{\bf l},\tau)\rangle
\nonumber\\
S({\bf k})&=&{1\over N}\sum_{{\bf j}\,\,{\bf l}}e^{i{\bf k}\cdot{\bf l}\,}
\langle \,\,c({\bf l})\,\,\rangle.
\label{eq:eq31}
\end{eqnarray}
Our normalization of the structure factor is such that if $c({\bf l})$
exhibits long range order, $S({\bf k_{*}})$ will be proportional to the
lattice volume $N=L_{{\rm x}}^{2}$, where $L_{{\rm x}}$ is
the linear extent in the spatial dimension.  If $c({\bf l})$ exhibits only
short range order, $S({\bf k_{*}})$ will be lattice size independent.
Here ${\bf k_{*}}=(\pi,\pi),(0,\pi),(\pi,0)$ are the possible ordering
wavevectors of the solid phase.

At weak coupling or high temperatures, $c({\bf l})$ exhibits only
short range order.  For ${\bf l}$ small, $c({\bf l})$ is enhanced but
very rapidly decays to its uncorrelated value $\rho^{2}$.  However, at
low temperatures for sufficiently large interactions, the density--density
correlations show long range oscillations.
The associated structure factor
$S({\bf k})$ evolves from being rather featureless to exhibiting a sharp peak
at ${\bf k_{*}}=(\pi,\pi)$ as $V_{1}$ increases, and a peak at
${\bf k_{*}}=(0,\pi)$ or $(\pi,0)$ as $V_{2}$ increases.  For our
2D system, for sufficiently large $V_{1}$, we expect a transition in the Ising
universality class. That is, $T_{c}$ is finite. In fact, if $t=0$ we have
$T_{c}=0.567\,\,V_{1}$. But even for a zero temperature phase
transition such as would occur at the Heisenberg point of the hard core
model, one will still observe ``long range order'' at finite
$T$ when the diverging
correlation length exceeds the spatial lattice size as $T$ is lowered.
In such instances, of course, a careful study of finite size effects is
required to draw conclusions concerning the existence of long range order.
Here we always report results for temperatures such that $\xi >L_{{\rm x}}$
so that observables have taken on their ground state values.
We have checked the scaling behavior to be sure that the ground
state is genuinely ordered, when so claimed.

Fig.~7 shows the superfluid density $\rho_{s}$ and structure factor $S(\pi,
\pi)$ as a function of $V_{1}$ for
$V_{0}=7$ and $V_{2}=0$. We see that at $V_{1}\approx 2.5$ there is a
phase transition from a superfluid to a solid phase.  The
transition on the $8\times 8$ lattice shown is already rather sharp; finite
size rounding in
the raw data for the structure factor and superfluid density near the
transition point is further reduced as one goes to $10\times 10$ lattices.
That one has true diagonal long range order in the solid phase is confirmed by
the fact that the structure factor scales linearly with the lattice volume.
Indeed, at $V_{1}=8$, $S(\pi,\pi)$ is almost precisely $100/64$ times
as large on the $10\times 10$ lattice than the $8 \times 8$.
There does not appear to be any window of coexistence between the superfluid
and solid phases at half--filling.
To within limits set by rounding, the transition points for
$S$ and $\rho_{s}$ coincide almost precisely.
We can make this statement more quantitative by performing
the appropriate scaling analysis on the data.  For example,
we have plotted $L_{{\rm x}}^{a}S(\pi,\pi)$ and $L_{{\rm x}}^{b}\rho_{s}$
versus $V_{1}$ for different
values of the exponent ratios $a,b$. Curves for different lattice sizes should
cross at the same critical value of $V_{1}$ for the appropriate choices of
$a,b$.  A complication is that the imaginary time lattice size
must be scaled as the appropriate power of the spatial
extent, and the dynamic exponent $z$ could be different
for the two transitions.  Making the simplest assumption
that $z$ is the same, however,
as was already suggested by the raw data, this scaling procedure shows
that the transition points for the two observables are within $0.5\%$ of each
other. While the structure factors do indeed cross nicely, the superfluid
density curves come together rather than pass through each other. This
seems to be a rather generic feature of simulations of the Bose Hubbard
model\cite{TRIVEDI} as opposed to related conserved current models
\cite{YOUNG,OTTERLO}.

As $V_{2}$ is increased, $c({\bf l})$
shows a similar transition from featureless uncorrelated behavior
to long range order, although in this case
$V_{2}$ favors the formation of a ``striped'' collinear phase
with alternating lines of occupied and empty sites. The structure factor
$S({\bf k})$ develops a peak at ${\bf k_{*}}=(\pi,0)$ or $(0,\pi)$.

In order to determine whether $V_{2}$ can drive a supersolid phase
at half--filling, we turn on
$V_{2}$ close to the point where the transition between superfluid and solid
occurs in Fig.~7. The density $\rho=0.5$. We show in Fig.~8 a plot of
$\rho_{s}$ and $S({\bf k})$ for ${\bf k}=(0,\pi),(\pi,0)$ and $(\pi,\pi)$. We
see that $V_{2}$ drives the N\'eel solid into a superfluid, and then at yet
larger values causes the formation of a striped solid phase. Again, the plots
suggest that there is no supersolid phase at $\rho=0.5$. Scaling plots similar
to those constructed at
$V_{2}=0$ do not reveal any evidence for distinct critical
points for superfluid and solid transitions to within our numerical accuracy.

We can put data from Figs.~7 and 8 together with similar runs for different
sweeps of $V_{1}$ and $V_{2}$ to obtain the ground state phase diagram of the
soft core BH model
at $V_{0}=7$ and $\rho=0.5$. This is shown in Fig.~9. At weak
couplings we have a superfluid phase, while at strong couplings there are two
possible solids: checkerboard and striped.  A strong coupling analysis
predicts a phase boundary between the solid phases at
$V_{2}=\frac12 V_{1}$
The superfluid phase extends out along this line in
a very robust manner, as opposed to the situation in
1D, where the superfluid window was rather narrow.\cite{NIYAZ}
This is a consequence of the highly degenerate nature of the
strong coupling ($t=0$) ground state along the line
$V_{1}=2V_{2}$. As can easily be seen, not only do the N\'eel and
checkerboard solids have the same energy, but an infinite number of defect
states are degenerate as well for $V_1=V_2$. For example in a horizontally
aligned collinear solid a whole column can be shifted up and down without
energy cost. \cite{FREERICKS,BRANDT} This large degeneracy stabilizes
superfluidity, even at large coupling.
We will comment further on this point when discussing the hard-core
phase diagram.

\noindent
{\bf Results off half--filling}

Although it does not appear that the BH model exhibits a supersolid phase at
$\rho=0.5$, we can see the coexistence of diagonal and off--diagonal long range
order when the filling is shifted away from $\rho=0.5$. In Fig.~10 we show
$\rho_{s}$ and $S(\pi,\pi)$ for the same parameters as Fig.~7 except now
$\rho=0.53$. We see that although $\rho_{s}$ declines significantly when the
solid forms, the excess boson density $\delta =\rho-0.5$ (the
magnetization $m$ is spin language) remains mobile in the
solid background. Indeed, simulations at different densities
(we found supersolids out to dopings of 0.675) show that the tail
in $\rho_{s}$ is precisely proportional to $\delta$. Fig.~11 shows the
analogous plot for a striped supersolid.  Note that we have here
separately displayed $\rho_{sx}$ and $\rho_{sy}$. As expected, the superfluid
density in the $x$ and $y$ direction is correlated with the direction in which
the striped solid channels run, as determined by the ordering wavevector
${\bf k_{*}}=(\pi,0)$ or $(0,\pi)$. If we had separately measured $\rho_{sa}$
and $\rho_{sb}$ on the two sublattices of the checkerboard solid, we would have
found an analogous symmetry breaking. The nonzero value of $\rho_{sa}-
\rho_{sb}$ is closely related to the appearance\cite{FISHERLIU} of a nonzero
order parameter $m_{xa}-m_{xb}$ in the language of the spin Hamiltonian Eq.~2.

If we were to use finite size scaling techniques
to locate the precise phase boundaries,
it would be necessary to scale the imaginary time length $L_{\tau}$
as a power of the spatial length $L_{{\rm x}}^{z}$, where $z$
is the dynamic critical exponent.  As we have earlier described,
it may be that different values of $z$ are associated with the
two transitions off half--filling, in which case the finite size scaling
analysis is much more delicate.\cite{OTTERLO,BATROUNI1}
We do not see the necessity of such a study here, since the
supersolid phase occupies an extended portion of the phase
diagram, and its existence is not predicated on proving the distinctness
of two transition points.

{\it Figures 10 and 11 provide compelling
evidence for the existence of a supersolid phase.
Our physical picture of this supersolid is one in which $\rho=0.5$ of the
bosons freeze into a rigid solid structure, while the remaining $\delta$
remain mobile. As we have seen, a signal of long range order then is present in
both the diagonal $\langle n_{i}n_{j}\rangle$ and off--diagonal $\langle
a_{i}a_{j}^{\dagger}\rangle$ channels.}

We have conducted our simulations of the BH Hamiltonian in the canonical
ensemble, and have presented our results by specifying the density $\rho$
rather than the chemical potential $\mu$. In describing the nature of the phase
diagram it is important to note that due to the existence of a gap in the solid
phases, the $\mu$--$\rho$ relation is non--trivial.  If the
gap is nonzero, when we dope our system
even slightly away from half--filling, the chemical potential is shifted
by a considerable amount. In the language of the spin Hamiltonian, Eq.~2, a
sizeable field $H_{z}$ is required to change the magnetization of the gapped
Ising phase. In Fig.~12 we illustrate this point by drawing the
$\mu/V_{1}$--$1/V_{1}$
phase diagram. A sweep at constant chemical potential reveals
a supersolid {\it window}. A sweep at fixed density skirts the pure solid and
remains in the supersolid phase. This is why we see in Figs.~10 and 11 a
supersolid for an extended region $V>V_{{\rm crit}}$ rather than in some narrow
region between phases exhibiting a single type of order.

If we now examine densities $\rho<0.5$, we find qualitatively
similar results: a superfluid phase gives way to a striped supersolid phase as
$V_{1}$ increases.  By these measures,
hole or particle doping appears qualitatively similar.
The same is true of the checkerboard supersolid, where
results for hole doping are entirely reminiscent
of the analogous particle doped case.

In fact, however, something rather different does go on
with particle and hole doping.
In Fig.~13 we show the ground state energy as a function of doping
for $V_{0}=7$, $V_{1}=3$, and $V_{2}=3$.
For these parameters, as
we have seen, we have a striped supersolid off half--filling and
a striped solid at $\rho=1/2$.  The change in slope of
$E_{0}$ at $\rho=1/2$ reflects a jump in the chemical potential which is,
in fact, just the gap in the solid phase.\cite{BATROUNI2}.
There is nothing particularly unusual here.
The strange feature occurs for the checkerboard case.
In Fig.~14 we show the ground state energy as a function of doping
for $V_{0}=7$, $V_{1}=3$, and $V_{2}=0$.
The fact that $E_{0}(\rho)$ is concave down for $\rho
<0.5$ indicates an instability to phase separation. Previous
studies\cite{PHASESEP} have suggested the possibility of phase separation in
systems with attractive boson interactions. However, we do not
have these Lennard--Jones type potentials here, only purely repulsive ones. It
is not immediately apparent
why mobile holes (or particles) in a rigid solid background should segregate
themselves.

A possible explanation, however, is as follows: Consider an isolated doped
hole in a checkerboard solid.
In order to move to another site of the same sublattice, it
must pass through an intermediate site on the opposite sublattice, a state of
energy $2V_{1}$. Thus the hole's effective hybridization is $t_{{\rm eff}}=
t^{2}/ 2V_{1}$. (This sort of argument has previously been used to predict the
shape of the phase boundary in the one dimensional extended BH Hamiltonian, in
good agreement with simulations.\cite{NIYAZ})  If two holes are near each
other, the intermediate state is lower in energy, so the effective
hybridization is increased. This suggests a possible mechanism for phase
separation: increased mobility of holes which propagate coherently. Of course,
the increase in $t_{{\rm eff}}$ is partially offset by the entropy cost of
confining one hole near the other.  Unfortunately, there appears to be an
analogous increase in $t_{{\rm eff}}$ for doped particles
which are proximate, so this reasoning does not explain
the fact that $E(\rho)$ is concave down for $\rho<0.5$ only.
Nevertheless, the simulations provide compelling evidence for
a lack of particle-hole symmetry.

In principle, one can also examine the issue of phase separation
through anomalies in $S({\bf k})$ for small
${\bf k}$.  However, our use of the canonical ensemble makes this
approach non-trivial.
Further work on the question of phase separation is needed.

\section{Simulations of the Hard Core Model}

We now examine the phase diagram in the hard core case.  This is important to
do for a number of reasons.  First, it allows us to make a connection
to the spin model limit, Eq.~2.  Second, as we have seen at $V_{0}=7$,
some of the interesting transitions occur at $V_{1}$ and $V_{2}$ values which
are getting rather large, while we expect in most physical
situations that
the on--site $V_{0}$ should be substantially greater than the near
neighbor interactions.  One consequence of this, is
that the doped bosons in the supersolid phase for our soft--core model
could move on the occupied sublattice, since the cost of
$V_{0}$ was less than the coordination number $z$ times the
near neighbor interaction strengths.  In the hard core model
such multiple occupancies are forbidden, and we want to make sure that
our conclusions are not affected by this change.

Fig.~15 shows results for the superfluid density and structure
factors for the half--filled case.  We sweep $V_{2}$ at fixed $V_{1}=
3$.  A N\'eel phase appears at small $V_{2}$.
For larger $V_{2}$ the superfluid phase appears
before making a transition into a collinear solid for yet larger $V_{2}$.
If $V_{1}$ is sufficiently small, the N\'eel phase at weak $V_{2}$
is eliminated, and the system remains superfluid down to
$V_{2}=0$.  Data for this and other sweeps is summarized in
Fig.~16 where the resulting ground state phase diagram is shown.
Note that we find the superfluid--N\'eel solid transition
at $V_{2}=0$ occurs at a value $V_{1}$ close to $2t$, which
is the result expected based on the mapping to the
spin model, Eq.~2.

As in the soft core case the weak coupling superfluid
extends out along the $V_{2}=V_{1}/2$ strong coupling boundary
between the two solid phases.  Unlike analogous studies\cite{NIYAZ} in
1D, this superfluid wedge is difficult to close, a phenomenon
which we earlier explained by the large degeneracy of competing
solid phases along the strong coupling line.
We have conducted simulations along the line $V_{2}=V_{1}/2$
and find that the superfluid density vanishes at $V_{1} \approx 7$.
Interestingly, there is no inset of solid order at this point.
This needs further study, for example to understand if some disordered
dimer phase might exist in this regime, in analogy with related
spin systems.

Fig.~17 is a plot for a doped lattice with
$\delta=\rho-1/2= 0.0625$.  The main difference is that, as in the
soft--core case, there is a superfluid tail after the structure factor
exhibits the transition into the solid phase.  That is, there
is a supersolid in the hard--core case as well.
As expected, doping inhibits somewhat the formation of
crystalline order, so that stronger couplings are required to
induce the crystalline order as is seen by comparing the
doped phase diagram, Fig.~18, with Fig.~16, the phase diagram
at half--filling.  Despite considerable rounding of the transitions,
scaling analyses conclude that the regions where $S$ is large
are indeed ordered.

Finally,
Figs.~19a,b show the ground state energy as a function of filling for
the hard--core model at $V_{1}=8, V_{2}=0$ (N\'eel solid), and
$V_{1}=8, V_{2}=4$ (collinear solid), respectively.
The data are qualitatively similar to the soft--core model.
In the collinear solid $E_{0}$ is concave up, with a change
in slope at $\rho=1/2$ which is the gap.  In the N\'eel solid
$E_{0}$ shows a tendency for phase separation.

\section{Related issues}

Up to now we have focussed on the ground state phase diagram of the BH
Hamiltonian. It is interesting to consider also the behavior of the system at
finite temperatures. Here the motion of doped bosons in the BH model which we
have studied with our simulations has a close connection with the idea of
``defectons'' in a solid\cite{ANDREEV1} where quantum tunneling caused by the
finiteness of the de Boer parameter delocalizes lattice defects at low
temperature. It is also of interest to study the behavior of the diffusion
constant $D$ for the full range of temperature. Here we expect that defects are
localized at high $T$, and $D$ first decreases exponentially as $T$ is lowered
in this classical regime. $D$ should then exhibit a plateau as quantum
diffusion takes over, and ultimately increase again as delocalization occurs.
While they focus largely on the behavior of single defectons, Andreev and
Lifshitz also consider the possibility of long range Coulomb interactions
causing localization into a ``defecton superlattice''. Our insulating
checkerboard solid is in fact an illustration of this. The Bose--Hubbard model
with only on--site $V_{0}$, has no solid phase at $\rho=0.5$, but
when $V_{1}$ is turned on, an ordered lattice does form.

We have focussed here on zero temperature,
the finite temperature phase diagram of the 2D BH model
would be interesting to study as well.  The solid transitions
are in the Ising universality class, and hence have a finite $T_{c}$.
Similarly one expects a Kosterlitz--Thouless type finite
critical temperature for the superfluid transition.
As for the topology of the phase diagram, several possibilities
have been explored by Liu and Fisher\cite{FISHERLIU}.
One intriguing case is the appearence of a tetracritical point;
where the three ordered phases (superfluid, supersolid and solid)
come together, giving way to the disordered phase with further
increase of the temperature.
This happens within a limited, but finite range of parameters on the mean field
level. The corresponding scaling theory was developed by Nelson and Fisher
\cite{NELSON}. Other alternatives include a supersolid phase
which exists only at finite temperatures, and that the tetracritical point
is split into bicritical points.\cite{FISHERLIU}
We hope to take up some of the issues in a further publication.


In the path integral representation of the BH partition function used in our
simulations, the particle number conservation leads to boson ``world lines''
propagating in the original 2 spatial dimensions plus an additional imaginary
time direction which runs from 0 to $\beta$. This picture has been used to
suggest close analogies between the physics of vortices in Type II
superconductors\cite{FISHER2} and the phase diagram of the 2D Bose--Hubbard
Hamiltonian. Frey, Nelson, and Fisher,\cite{FREY} have recently discussed both
thermally driven and quantum phase transitions, for example as caused by the
introduction of defects or interstitials into the Abrikosov lattice, in these
vortex systems. This also has close connections with the results we have
discussed here.

\section{conclusions}

In this paper we have
considered quantum phase transitions in the Bose--Hubbard
hamiltonian. We identified several phases: solid and supersolid phases with
N\'eel and collinear patterns, furthermore a superfluid and a Mott type
insulating phase. The phase diagram has been determined analytically
and the spin--wave spectrum has been calculated. The dynamical critical
exponents at each transitions were calculated and preexisting controversies
were settled. Our numerical work -- utilizing Quantum Monte Carlo methods --
provided a detailed study of the different phases. Concerning the phase diagram
the existence of supersolid phases has been forcefully confirmed.
These phases exist only off half--filling, in accordance with the mean field
results, but in disagreement with some recent claims.
The possibility of phase separation in the model has been investigated as well,
and provides evidence for a violation of the previously assumed
particle--hole symmetry of the model.

\centerline{{\bf Acknowledgements}}
We acknowledge useful discussions with D. Arovas. This work was supported by
the
National Science Foundation grant DMR 92-06023,
and by Los Alamos National Laboratory under LACOR grant No.
UC-94-2-A-213. APK gratefully acknowledges
support through a habilitation scholarship of the Deutsche
Forschungsgemeinschaft (DFG). GTZ acknowledges the kind hospitality
at the University of Karlsruhe, where his work was supported by the
Sonderforschungsbereich 195 of the DFG.
Much of the numerical work was performed on a Connection Machine 5
at Thinking Machines Corporation.

\newpage

\noindent
\centerline{{\bf Figure Captions}}
\vskip0.5cm
\noindent
{\bf Fig.~1:} Mean field phases MF of the XXZ spin
Hamiltonian Eq.~2 on a 2D square lattice.

\noindent
{\bf Fig.~2:} Hard core mean field phase diagram at half--filling
$\rho={1\over 2}$ from comparing the energies of superfluid, N\'eel and
collinear solid.

\noindent
{\bf Fig.~3:} Hard core mean field phase diagram (bold lines) away from
half--filling for $m=|\rho-{1\over 2}|=0.2$ from comparing the energies of
superfluid, N\'eel and collinear supersolid. Thin lines indicate the phase
boundaries at half--filling $m=0$ (see Fig. 2).

\noindent
{\bf Fig.~4:} Hard core mean field phase diagram, magnetisation $m$ versus
$V_2$, for fixed $V_1=3$. SS denotes the supersolid phases.

\noindent
{\bf Fig.~5:} Hard core mean field phase diagram, magnetisation $m$ versus
$V_1$, for fixed $V_2=1$.

\noindent
{\bf Fig.~6:} Spin wave dispersions in the (a) N\'eel solid, (b) at the N\'eel
solid--N\'eel supersolid boundary, (c) in the N\'eel supersolid, (d) at the
N\'eel supersolid--superfluid boundary, and (e) in the superfluid. In all plots
$V_2$ and $V_1$ are fixed to $V_2=1.5$, $V_1=4.$ and the magnetic field $h$
($H_z$ in the text) is varied.

\noindent
{\bf Fig.~7:} The superfluid density $\rho_{s}$ and $S(\pi,\pi)$ as a function
of $V_{1}$ for $V_{0}=7$ and $V_{2}=0$. The density $\rho=0.5$ and $\beta=4$.
The transitions in $\rho_{s}$ and $S(\pi,\pi)$ appear to occur at roughly the
same value of $V_{1}$.

\noindent
{\bf Fig.~8:} The superfluid density and structure factor as a function of
$V_{2}$ at $V_{0}=7$, $V_{1}=2.75$, and $\rho=0.5$.

\noindent
{\bf Fig.~9:} The ground state phase diagram of the BH model at $\rho=0.5$ and
with a soft core on--site repulsion $V_{0}=7$.

\noindent
{\bf Fig.~10:} The superfluid density and structure factor for the same
parameters as in Fig.~7, except now the system is doped to $\rho=0.53$. A
superfluid tail remains in the (checkerboard) solid phase.

\noindent
{\bf Fig.~11:} The superfluid density and structure factor for the same
parameters as in Fig.~8, except now the system is doped to $\rho=0.56$. A
superfluid tail remains in the (striped) solid phase.

\noindent
{\bf Fig.~12:} The qualitative $T=0$ phase diagram of the Bose--Hubbard model
is illustrated. A sweep at constant $\mu$ (full arrow) could cut across the
phase boundaries as shown, revealing a supersolid window, while a sweep of
constant density (dashed arrow) remains in the supersolid phase at strong
coupling. The Mott insulating phase at large $\mu$ has density one boson per
site.

\noindent
{\bf Fig.~13:} The ground state energy as a function of density for
$V_{0}=7$, $V_{1}=3$, $V_{2}=3$.

\noindent
{\bf Fig.~14:} The ground state energy as a function of density
for $V_{0}=7$, $V_{1}=3$, $V_{2}=0$.

\noindent
{\bf Fig.~15:}  Superfluid density and structure factors versus
$V_{2}$ for $V_{1}=3$.  The density $\rho=0.500.$

\noindent
{\bf Fig.~16:}  The phase diagram of the half--filled hard--core model.
The dashed lines are the results of the Mean Field analysis presented
earlier in the paper.

\noindent
{\bf Fig.~17:}  Superfluid density and structure factors versus
$V_{2}$ for $V_{1}=3$.  The density $\rho=0.563.$

\noindent
{\bf Fig.~18:}  The phase diagram of the doped hard--core model.

\noindent
{\bf Fig.~19:}  Ground state energy versus density for the hardcore
model.  (a) $V_{1}=4, V_{2}=0$ and (b) $V_{1}=4, V_{2}=4$.

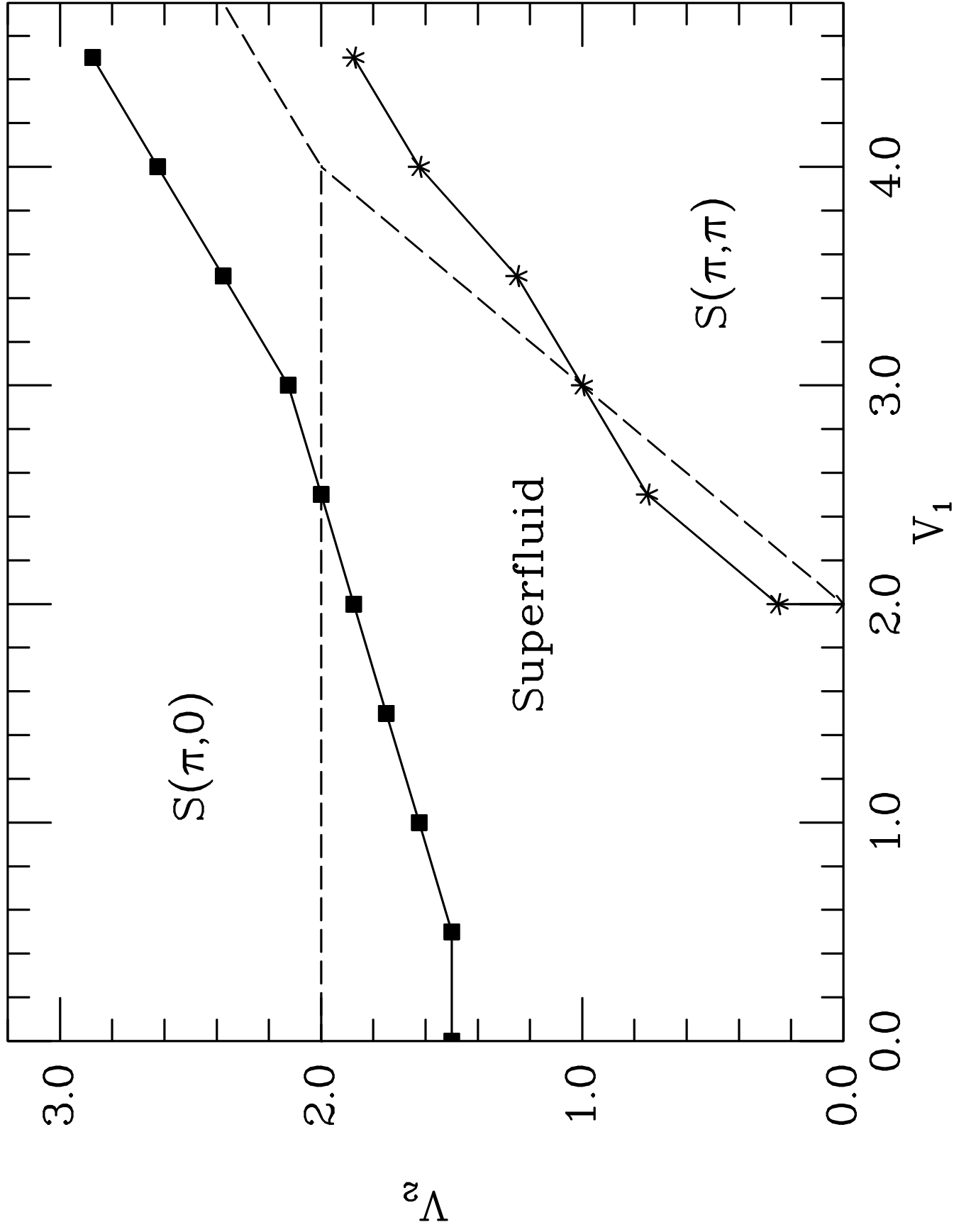
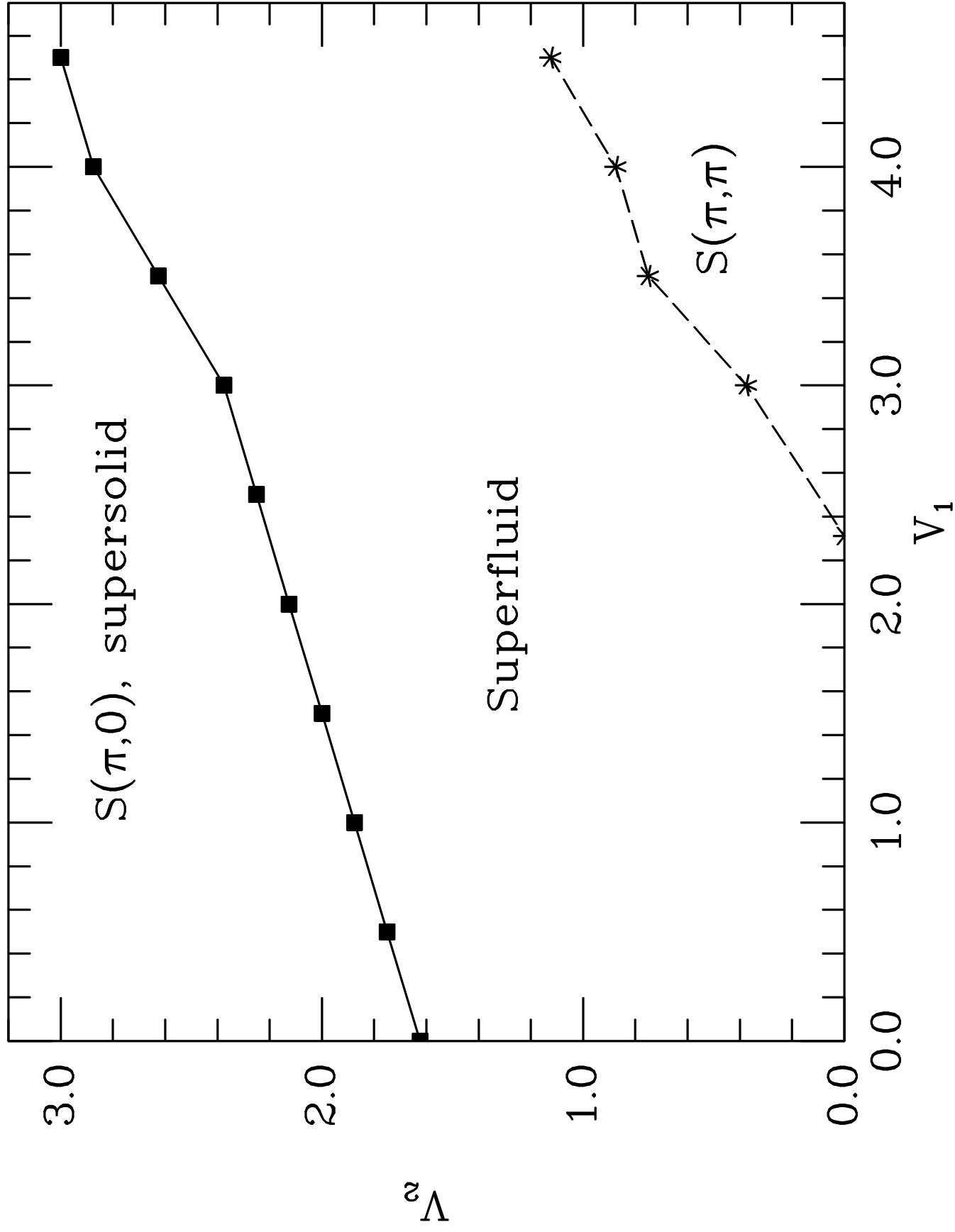
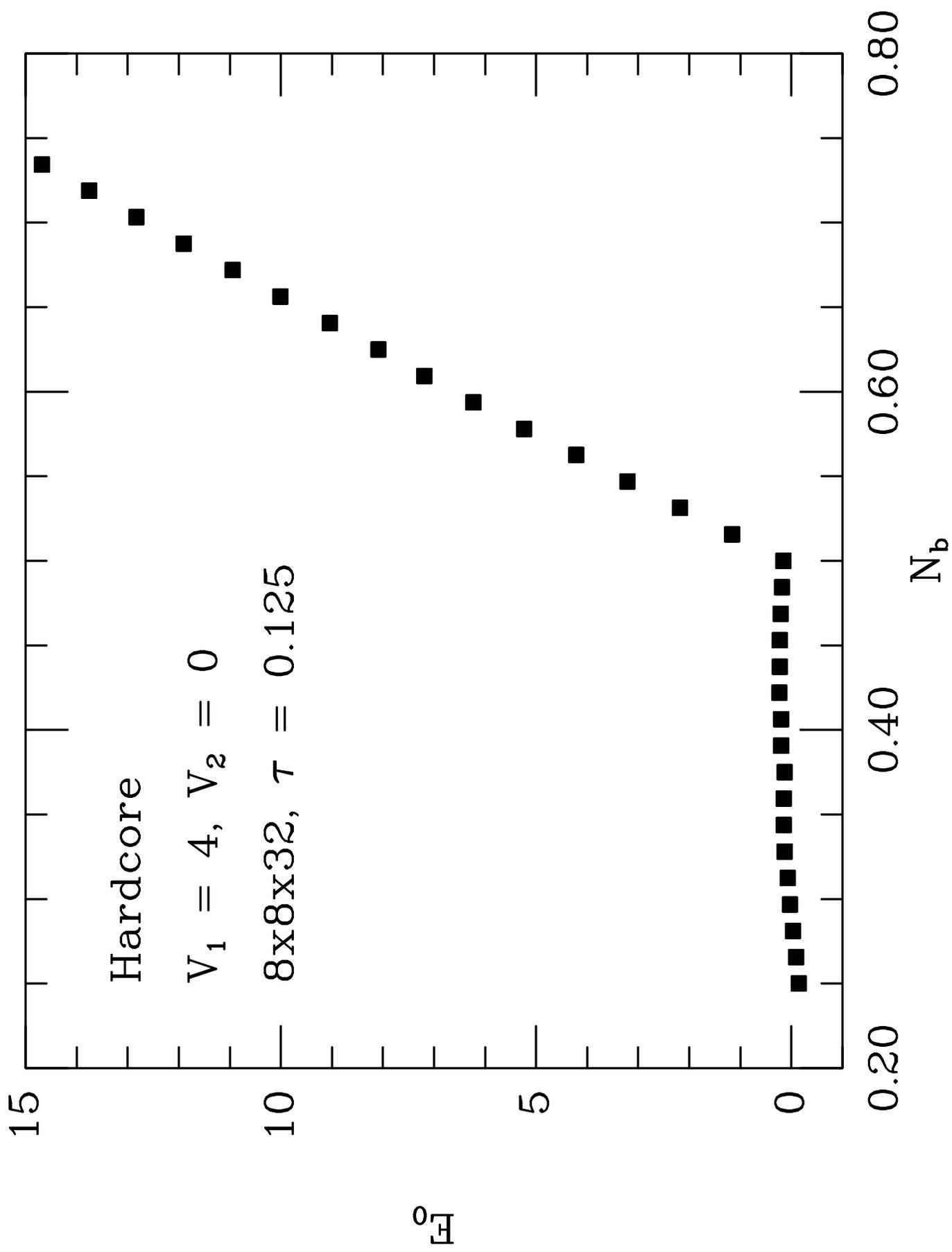
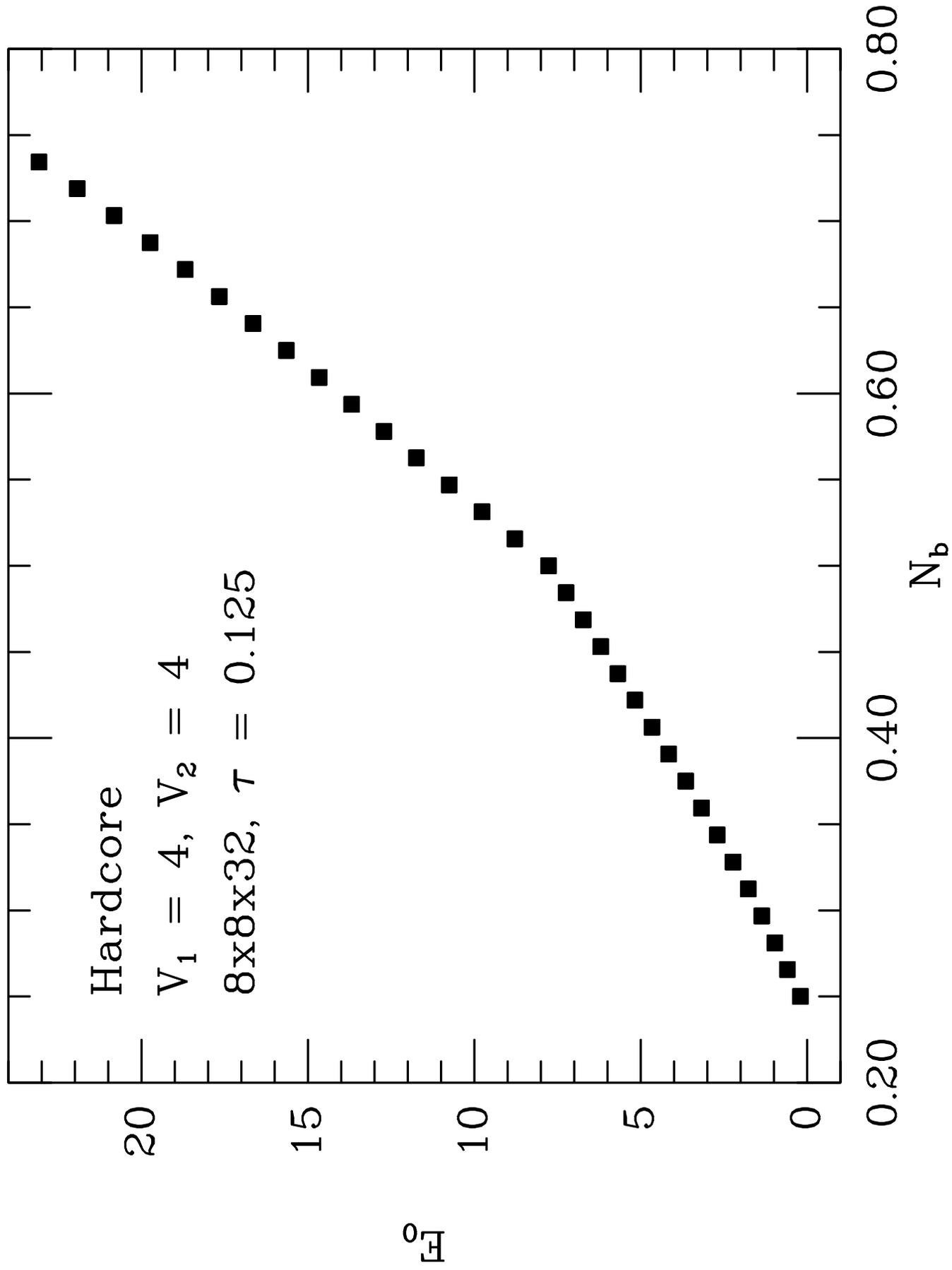
\end{document}